# Nonlinear imaging of nanoscale topological corner states

Sergey S. Kruk[1,2]*, Wenlong Gao[1,2], Duk-Yong Choi[3], Thomas Zentgraf[2], Shuang Zhang[4,5,6], and Yuri Kivshar[1]*

[1]Nonlinear Physics Center, Research School of Physics, Australian National University, Canberra ACT 2601, Australia
[2] Department of Physics, Paderborn University, 33098 Paderborn, Germany
[3]Laser Physics Center, Research School of Physics, Australian National University, Canberra ACT 2601, Australia
[4]School of Physics and Astronomy, University of Birmingham, Birmingham B15 2TT UK.
[5]Department of Physics, University of Hong Kong, Hong Kong, China
[6]Department of Electrical & Electronic Engineering, University of Hong Kong, Hong Kong, China

*sergey.kruk@outlook.com *yuri.kivshar@anu.edu.au

## ABSTRACT

Topological states of light represent counterintuitive optical modes localized at boundaries of finite-size optical structures that originate from the properties of the bulk. Being defined by bulk properties, such boundary states are insensitive to certain types of perturbations, thus naturally enhancing robustness of photonic circuitries. Conventionally, the N-dimensional bulk modes correspond to (N–1)-dimensional boundary states. The higher-order bulk-boundary correspondence relates N-dimensional bulk to boundary states with dimensionality reduced by more than 1. A special interest lies in miniaturization of such higher-order topological states to the nanoscale. Here, we realize nanoscale *topological corner states* in metasurfaces with $C_6$-symmetric honeycomb lattices. We directly observe nanoscale topology-empowered edge and corner localizations of light and enhancement of light–matter interactions via a nonlinear imaging technique. Control of light at the nanoscale empowered by topology may facilitate miniaturization and on-chip integration of classical and quantum photonic devices.

Topological states possessing quantized invariants have been a subject of intense studies in various fields ranging from condensed matter physics to photonics, due to a plethora of novel physics and potential applications [1],[2]. Topological photonic systems have been applied to lasing [3]– [6], quantum computing platforms [7]–[9], excitons [10], exciton-polaritons [11], and robust signal transmission [12]. Principles of topological photonics are of a particular interest for control of light at the nanoscale. A promising pathway towards nanophotonic topological states may rely on judiciously designed all-dielectric subwavelength elements hosting electric and magnetic Mie multipoles. Nanostructures assembled from such resonators have been used to demonstrate the simplest types of topological *first-order bulk-boundary correspondence*, such as one-dimensional (1D) states in two-dimensional (2D) structures [12], [41] and zero-dimensional (0D) states existing in 1D structures [42]–[44].

The *higher-order bulk-boundary correspondence* in topologically nontrivial structures, including 0D topological *corner states* in 2D structures, has recently come into focus throughout various branches of physics including condensed matter physics [13]–[20], electrical circuit systems [21], phononics and mechanics [22], [23], acoustics [24]–[27], and microwaves [28]–[33]. Topological nature of such corner states has been suggested to lead to their robustness against various types of disorder [34]. In optics, topological corner states have been first realized in arrays of waveguides [35], [36] and in coupled ring resonators [37]. However, both types of systems operate with constituent elements that are larger than the wavelength of light by an order of magnitude. Photonic crystals representing a 2D analog of 1D Su–Schrieffer–Heeger (SSH) systems have been demonstrated to support corner localizations of light at a smaller scale [38-40].

In this work, we report 0D corner states in a 2D topological nanostructure based on a photonic analog of the spin-Hall effect. We observe directly 2D bulk modes, 1D waveguiding edge states and 0D localized corner states of light. In our studies, we map nonlinear enhancement of the third-harmonic signal generated by the topology-empowered light localizations. A shorter wavelength of the third harmonic signal reduces the diffraction limit of the far-field imaging, therefore yielding high-resolution background-free imaging of topological states.

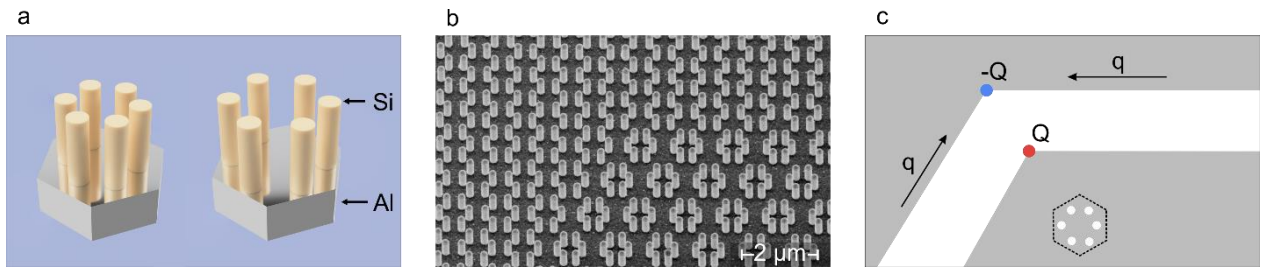

Figure 1. Hybrid metal–dielectric metasurfaces supporting nanoscale topological corner states of light. (a) Illustration of two types of constituent elements: shrunken (left) and expanded (right) hexamers made of silicon pillars on a metal substrate. (b) Scanning electron micrograph of the fabricated metasurface sample consisting of two topologically dissimilar domains formed by the shrunken and the expanded hexamers. (c) Schematics of two localized states at 120° and 240° corners at the domain wall between the shrunken (white) and the expanded (gray) hexamer metasurfaces. Q denotes the corner charge, and q is the edge polarization.

The metasurfaces consist of two types of unit cells: expanded and shrunken silicon hexamers (see Fig. 1a) hosting Mie resonances. The resonant response gives rise to nontrivial topology of the structure conceptually similar to an earlier theoretical proposal [2]. Our metasurfaces are designed on top of an aluminum substrate. The metal substrate reflects transverse magnetic (TM)-like modes effectively doubling the height of the pillars. At the same time, the transverse electric (TE)-like modes are suppressed [36]. This design is at the same time compatible with standard nanofabrication technologies. The band-folding geometry of this structures leads to nontrivial topology [45]. The energy spectrum of such metasurfaces features a band gap between its 2D bulk modes. 1D topological edge states reside in the bandgap of the bulk modes. The edge state spectrum, in its turn, features a mini-band-gap that arises due to symmetry reduction at the boundary. This is associated with higher-order topological effects [46]. 0D topological corner states reside within the mini-bandgap of the 1D states. The 0D corner states are associated with the nontrivial fractional corner charge localization [47].

Localized topological states arise at the corners between topologically dissimilar domains. Such corner localization of light in photonic systems can be associated with similar effects in solid-state physics where the nontrivial topology of the corner states was found recently to arise from the charge filling anomalies for electrons in crystals [47]. Specifically, at charge neutrality, the arrangements of electrons at the maximal Wyckoff positions in crystals inevitably break the intrinsic symmetries ($C_6$ symmetry in our design), and extra electrons or holes have to be added onto corners of the crystal to restore the crystalline symmetry. However, this causes the charge neutrality to be broken, and introduces a new corner charge term Q into the well-established expression: $Q_c=q_{n1}+q_{n2}$, and the original equation is modified to $Q_c=Q+q_{n1}+q_{n2}$, where $q_{n1}$ and $q_{n2}$ are the edge polarizations corresponding to the edges forming the corner and $Q_c$ is the total corner charge localization. This results in the 120° corner having topological protected e/2 fractional charge localization [47]. The 240° corner also naturally holds e/2 fractional charge of opposite sign because the 120° corner can be cut out from one infinite crystal with charge neutrality. The corner charge distribution for both the 120° and the 240° corner configurations is schematically illustrated by a ribbon configuration shown in Fig. 1c. We further note here that the discussed topological corner localization exists for corners formed by the domain walls with an armchair shape (from a perspective of the arrangement of individual pillars, otherwise a zig-zag shape from a perspective of the boundaries of unit cells. We will further use the perspective of individual pillars, and hence will refer to this type of domain wall as an "armchair" type).

We proceed with designing the metasurfaces using full-wave 3D simulations with COMSOL Multiphysics with details provided in Methods The final structure consists of Si cylinders 180 nm in diameter and 580 nm in height organized into hexamer clusters with a lattice constant of 1100 nm and shrunken/expanded coefficients of 0.764 and 1.12 correspondingly. The Si nanostructure rests on an Al mirror substrate.

We fabricate the designed structures from amorphous silicon using a standard electron beam lithography (see an electron microscopy image in Fig. 1b and details in the Methods). We directly observe on a camera in the far-field 2D bulk modes, 1D edge states and 0D corner states in our metasurfaces with a nonlinear imaging technique [6,7] (see details in Methods). In our experiments, a small portion of infrared optical energy is converted by the metasurfaces into visible light via a third-harmonic generation process. Specifically, we illuminate large areas of our sample with a short-pulse laser system in 1550-1750 nm spectral range and detect the third harmonic between 517 and 583 nm correspondingly. Third harmonic conversion process takes place within silicon hexamers due to intrinsic nonlinearities of the material. Nonlinear imaging approach provides multiple benefits. As the third-harmonic signal is no longer confined within the structure's modes, it radiates freely into the far-field carrying information about the near-field light distribution. In the detection system, the third-harmonic visible light can be easily filtered out from the infra-red excitation thus providing a background-free information. Nonlinear dependence of the third harmonic intensity on the excitation optical intensity enhances the contrast of nonlinear imaging compared to its linear counterpart. A shorter wavelength of third harmonic increases the resolution of diffraction-limited far-field imaging. Thus, nonlinear imaging is uniquely suitable for direct detailed visualizations of nanoscale optical modes, including subwavelength topological edge and corner states.

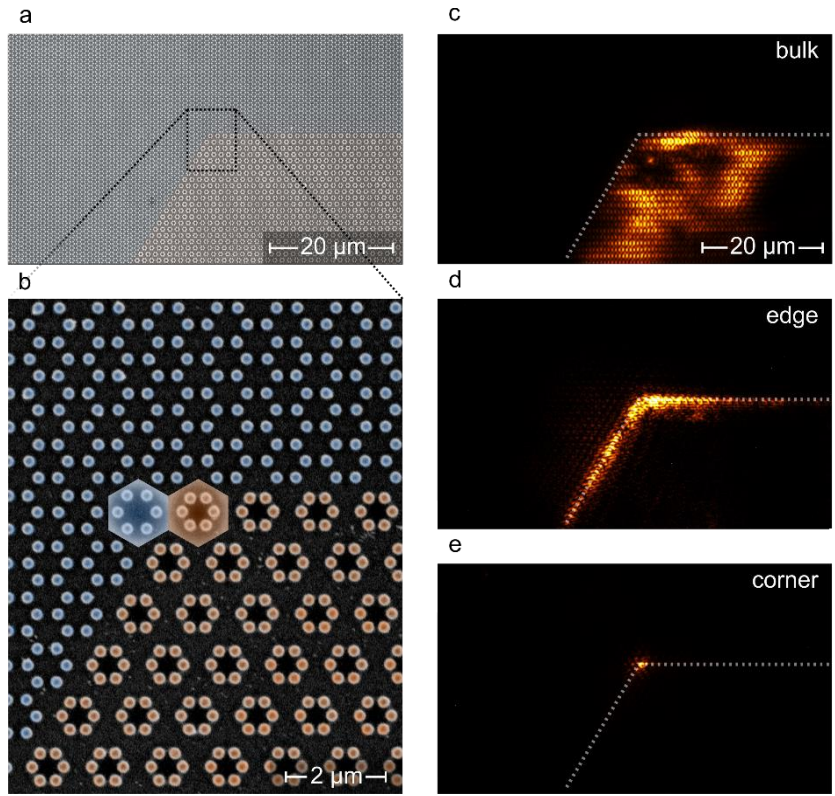

Figure 2. Experimental observation of topological light localizations in metasurfaces. (a) False-color scanning electron micrograph of the metasurface featuring a 240° corner between two topologically dissimilar areas. (b) Close-up image of the corner with the two marked unit-cells featuring the shrunken and expanded lattice designs. (c-e) Nonlinear optical microscopy photographs of the bulk mode at 1730 nm wavelength (c), the edge state at 1590 nm wavelength (d) and the corner states at 1615 nm wavelength (e). The images show a third-harmonic signal enhanced by the corresponding bulk, edge, and corner light localizations.

In our system of shrunken and expanded hexamers, the optical eigenstates are circularly polarized, as determined by the structure symmetry. Therefore, in our experiments we employ a right-circularly polarized incident beam to couple efficiently from the far-field to the edge states and to the corner states. We further scan the angles of incidence of the illumination beam within approximately a 40° cone (see details in Methods) to find empirically the wavevectors that are optimal for coupling to the edge and the corner states from the far-field.

Figures 2 a,b show false-color electron microscope images of a 240° corner angle formed by two armchair domain walls. Figures 2c-e show false-color camera photographs of 2D bulk modes, 1D edge states, and 0D corner states at their corresponding excitation wavelengths. The photographs were obtained at the third harmonic wavelengths with the nonlinear imaging technique.

We complement our studies of the 240° corner formed by the arm-chair domain walls (arm-chair corner) with experimental and theoretical analysis of a 120° arm-chair corner contrasted to a 240° corner formed by a zigzag domain wall (a zig-zag corner).

We calculate eigenmodes of three metasurfaces hosting the corresponding three types of corners using full-wave numerical simulations in COMSOL (see Figs. 3a-c). Figures 3a,b that correspond to metasurfaces with an armchair domain wall feature a small band-gap and a mid-gap state. In contrast, figure 3c corresponding to a zig-zag domain wall does not demonstrate a pronounced bandgap in the energy spectrum (band diagram can be found in Supporting Information Figure S1). Both the gaplessness of the edge state on the zig-zag domain, and the 120° and 240° corner states at the armchair domains can be understood from the structure of the one-dimensional edge Hamiltonian with alternating-sign coupling coefficient whose gap size is determined by the difference in on-site energies. (Supporting Information).

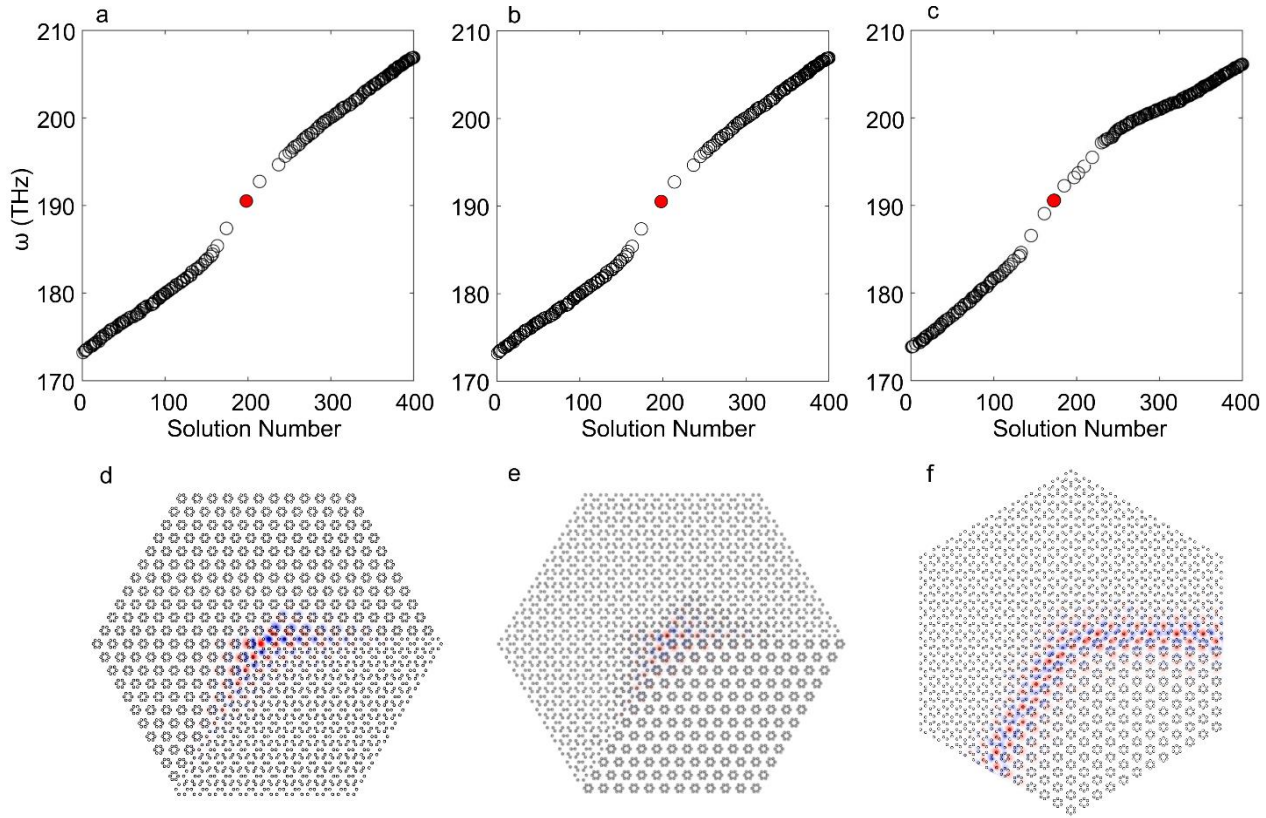

Figure 3. Theoretical studies of nanoscale topological corner states. (a-c) Eigensolutions spectra of the metasurfaces with (a) 120° arm-chair corner (b) 240° arm-chair corner. (c) 240° zig-zag corner. (d-f) Electric field distribution corresponding to eigensolutions marked with red dots in a-c. The 120° corner (d) features anti-symmetric electric field phase distribution, while 240° (e) shows symmetric field distribution with respect to the bisector of the corner.

Figures 3d-f show full-wave numerical calculations of electric field distributions for the three types of corners for energies corresponding to the mid-gap state position. We numerically verify robustness of the topological corner states against lattice disorder by adding perturbations onto the nano pillars' location in both x and y directions that satisfy normal distributions. It was found that under the same standard deviation, topological corner states are well localized. We additionally find a trivial corner localization and demonstrate that under same perturbation the trivial localization in striking contrast becomes poorly localized, mixing with other localizations on the edge and in the bulk (see details in Supporting Information).

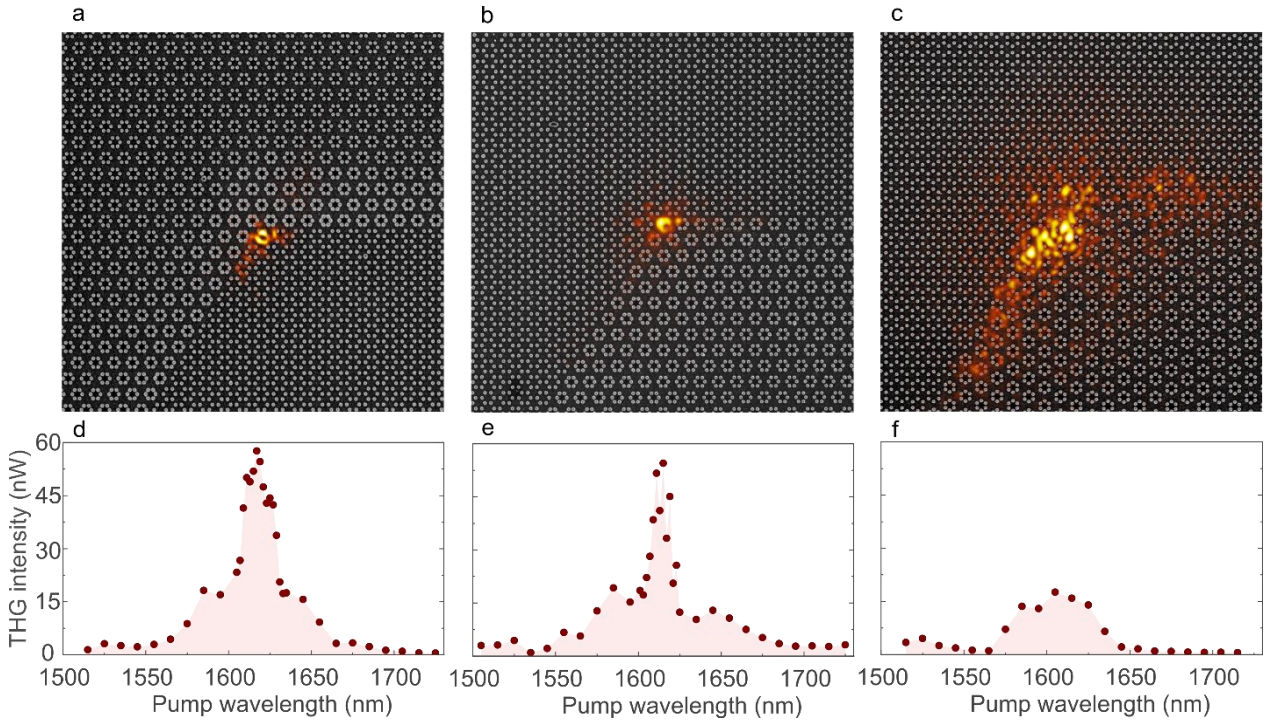

**Figure 4. Experimental characterization of the three types of corners.** (a-c) Superimposed electron microscope images visualizing detailed structures of the metasurfaces (gray) with visible camera false-color photographs depicting distribution of the third harmonic signal at same locations. All camera images were taken for 1615 nm excitation wavelength, thus corresponding to 538 nm third-harmonic wavelength. (a) 120° arm-chair corner. (b) 240° arm-chair corner. (c) 240° zig-zag corner. The electron microscope image background serves for illustrative purposes only as alignment with a camera photograph is performed with limited accuracy due to much lower resolution of the optical microscopy. (d-f) Peak intensity of the third-harmonic signal integrated over the area of the corner hexamer versus pump wavelength.

In our experiments we directly observe corner localizations of light for both types of topological corners (see Figs. 4a,b). The signal is largely confined within just a single corner hexamer. In this scenario, no significant coupling to the edge states along the domain walls was detected, as expect for structures exhibiting a corner state in the center of the mini-gap (as per Figs. 3a, b).

In a sharp contrast, the non-topological corner does not demonstrate a localized state (Fig. 4c). Light distribution in the vicinity of the non-topological corner spreads over multiple unit cells in a random fashion. The light tends to be distributed additionally along the domain wall, as in this case the structure does not feature a mini gap as can be seen in Fig. 3c.

Figures 4d-e show the enhancement of the nonlinear signal at the wavelength attributed to the corner state. The spectra were obtained by measuring the third harmonic intensity generated by the corner hexamer as the pump wavelength varied. All other pump parameters (e.g. polarization and incident angle) were kept unchanged. We estimate the ratio of the third harmonic peak power to pump peak power $P^{3\omega}_{peak}/P^{\omega}_{peak}$ as $1.8\times10^{-7}$ with details on excitation beam parameters and evaluation methods presented in the Supporting Information.

In this work, we have directly observed photonic topological corner states at the nanoscale in hybrid metal-dielectric metasurfaces. The Mie-resonant metasurfaces represented a photonics analogue of the spin-Hall effects. The corner states facilitated strong localizations of light at the same time providing topological protection against substantial perturbations of the system. We observed an enhancement of nonlinear light-matter interactions driven by topological light localizations and employed nonlinear imaging technique. We mapped 2D bulk modes, 1D edge states and 0D corner states in the far-field at the third harmonic frequency. Nonlinear imaging technique provided us with background-free, high-contrast, high-resolution photographs of the states. Topological control of light at the nanoscale and topologically-driven enhancement of light-matter interactions may facilitate the development of on-chip nanophotonic classical and quantum devices.

# SUPPORTING INFORMATION

Band diagrams of the metasurfaces. Note on the origin of gaplessness of the zig-zag edge states. Analysis of topological robustness against disorder. Estimations of nonlinear conversion efficiency.

# ACKNOWLEDGEMENTS

S.K. thanks Kirill Koshelev for an expert advice. Y.K. thanks Alexander Poddubny for encouraging discussions. W.G. is indebted to Daria Smirnova for her generous help and useful advice. S.K. acknowledges support from the Alexander von Humboldt Foundation. D.Y.C. acknowledges the use of the Australian National Fabrication Facility of the ACT Node. This work was also supported by the Australian Research Council (grant DP200101168) and the European Research Council (ERC) under the European Union's Horizon 2020 research and innovation program (grant agreements No. 724306 and 648783).

# MATERIALS AND METHODS

**Theoretical analysis.** We use COMSOL finite-element-method solver in frequency domain and the eigenmode solver. In our calculations metasurfaces are placed on a semi-infinite substrate.

The tight binding model is used to analyze the semi-analytical counterpart of the studied model. In the tight binding model, each silicon pillar was assumed to hold an s orbital wave function. By changing the on-site potential of three sites in the corner unit-cell. The corner states can be transferred throughout the entire bandgap, forming a gapless spectrum flow.

**Fabrication.** We deposit a layer of Al on Si substrate using an electron-beam evaporation (Temescal BJD-2000) followed by a deposition of hydrogenated amorphous Si film using plasma-enhanced chemical vapor deposition (Oxford Plasmalab System 100). The film thickness was was checked with the ellipsometer JA Woollam M-2000D. We next spin-coated an electron beam resist ZEP 520A (2:1) anisole. Next, the resist was patterned by an electron-beam lithography (Raith-150 EBL system, exposure dose 140 µA s/cm$^2$). A hard Al mask was created via metal deposition and a lift-off process. An inductively coupled plasma reactive ion etching was used to transfer the mask pattern onto the silicon layer (Oxford Plasmalab System 100 with $CHF_3$, $SF_6$ and Ar gases). The hard mask was subsequentially removed.

**Optical experiments.** The sample was excited with the light of a tunable optical parametric amplifier MIROPA from Hotlight Systems pumped by a pulsed laser Ekspla Femtolux. The excitation regime used 5.77 ps pulses with 5.144 MHz repetition rate and 300 mW of average output power at 1615 nm wavelength. Lasing spectra were controlled by a fiber-coupled spectrometer (NIR-Quest by Ocean Optics). Spectra of generated visible light were measured by a fiber-coupled spectrometer QE Pro by Ocean Optics confirming that the signal corresponds to a third-harmonic wavelength. Spatial distributions of visible light across the samples were imaged by a cooled camera (Trius-SX694 Starlight Xpress Ltd) paired with an infinity-corrected f=150 mm achromatic doublet lens from Thorlabs. To illuminate the sample, a collimated laser beam was narrowed by two lenses: a f=100 mm achromatic doublet from Thorlabs, and an objective lens ×100 0.7NA from Mitutoyo. The same objective lens was collecting the third-harmonic signal. A dichroic mirror was used to separate the pump beam and the backward emitting third harmonic signal. The polarization of the pump beam was controlled after the dichroic mirror to avoid its possible influence on the polarization state. In reflection, the imaging system residuals of the pump beam were filtered with an FGS900 filter. Variation of the sample illumination angle is performed via variation of a focal point of the f=100 mm lens within the back focal plane of the Mitutoyo objective.

# REFERENCES

[1]     L. Lu, J. D. Joannopoulos, and M. Soljačić, "Topological photonics," Nat. Photonics, vol. 8, no. 11, pp. 821–829, 2014.

[2]     Y. Liu, X. Chen, and Y. Xu, "Topological Phononics: From Fundamental Models to Real Materials," Adv. Funct. Mater., vol. 30, no. 8, p. 1904784, Feb. 2020, doi: 10.1002/adfm.201904784.

[3]     B. Bahari, A. Ndao, F. Vallini, A. El Amili, Y. Fainman, and B. Kanté, "Nonreciprocal lasing in topological cavities of arbitrary geometries.," Science, vol. 358, no. 6363, pp. 636–640, Nov. 2017, doi: 10.1126/science.aao4551.

[4]     M. A. Bandres et al., "Topological insulator laser: Experiments," Science, vol. 359, no. 6381, Mar. 2018, doi: 10.1126/science.aar4005.

[5]     P. St-Jean et al., "Lasing in topological edge states of a one-dimensional lattice," Nat. Photonics, vol. 11, no. 10, pp. 651–656, Oct. 2017, doi: 10.1038/s41566-017-0006-2.

[6]     Z. K. Shao et al., "A high-performance topological bulk laser based on band-inversion-induced reflection," Nat. Nanotechnol., vol. 15, no. 1, pp. 67–72, Jan. 2020, doi: 10.1038/s41565-019-0584-x.

[7]     S. Barik et al., "A topological quantum optics interface," Science, vol. 359, no. 6376, pp. 666–668, Feb. 2018, doi: 10.1126/science.aaq0327.

[8]     A. Blanco-Redondo, B. Bell, D. Oren, B. J. Eggleton, and M. Segev, "Topological protection of biphoton states," Science, vol. 362, no. 6414, pp. 568–571, Nov. 2018, doi: 10.1126/science.aau4296.

[9]     E. Strambini, S. D'ambrosio, F. Vischi, F. S. Bergeret, Y. V. Nazarov, and F. Giazotto, "The ω-SQUIPT as a tool to phase-engineer Josephson topological materials," Nat. Nanotechnol., vol. 11, no. 12, pp. 1055–1059, Dec. 2016, doi: 10.1038/nnano.2016.157.

[10]    D. Varsano, M. Palummo, E. Molinari, and M. Rontani, "A monolayer transition-metal dichalcogenide as a topological excitonic insulator," Nat. Nanotechnol., vol. 15, no. 5, pp. 367–372, May 2020, doi: 10.1038/s41565-020-0650-4.

[11]    S. Klembt et al., "Exciton-polariton topological insulator," Nature, vol. 562, no. 7728, pp. 552–556, Oct. 2018, doi: 10.1038/s41586-018-0601-5.

[12]    M. I. Shalaev, W. Walasik, A. Tsukernik, Y. Xu, and N. M. Litchinitser, "Robust topologically protected transport in photonic crystals at telecommunication wavelengths," Nature Nanotechnology, vol. 14, no. 1. Nature Publishing Group, pp. 31–34, Jan. 01, 2019, doi: 10.1038/s41565-018-0297-6.

[13]    W. A. Benalcazar, B. A. Bernevig, and T. L. Hughes, "Quantized electric multipole insulators," Science, vol. 357, no. 6346, pp. 61–66, Jul. 2017, doi: 10.1126/science.aah6442.

[14]    J. Langbehn, Y. Peng, L. Trifunovic, F. Von Oppen, and P. W. Brouwer, "Reflection-Symmetric Second-Order Topological Insulators and Superconductors," Phys. Rev. Lett., vol. 119, no. 24, p. 246401, Dec. 2017, doi: 10.1103/PhysRevLett.119.246401.

[15]    Z. Song, Z. Fang, and C. Fang, "(d-2) -Dimensional Edge States of Rotation Symmetry Protected Topological States," Phys. Rev. Lett., vol. 119, no. 24, p. 246402, Dec. 2017, doi: 10.1103/PhysRevLett.119.246402.

[16]    F. Schindler et al., "Higher-order topological insulators," Sci. Adv., vol. 4, no. 6, p. eaat0346, Jun. 2018, doi: 10.1126/sciadv.aat0346.

[17]    M. J. Park, Y. Kim, G. Y. Cho, and S. Bin Lee, "Higher-Order Topological Insulator in Twisted Bilayer Graphene," Phys. Rev. Lett., vol. 123, no. 21, p. 216803, Nov. 2019, doi: 10.1103/PhysRevLett.123.216803.

[18]    Z. Yan, F. Song, and Z. Wang, "Majorana Corner Modes in a High-Temperature Platform," Phys. Rev. Lett., vol. 121, no. 9, p. 096803, Aug. 2018, doi: 10.1103/PhysRevLett.121.096803.

[19]    Q. Wang, C. C. Liu, Y. M. Lu, and F. Zhang, "High-Temperature Majorana Corner States," Phys. Rev. Lett., vol. 121, no. 18, p. 186801, Oct. 2018, doi: 10.1103/PhysRevLett.121.186801.

[20]    C. H. Hsu, P. Stano, J. Klinovaja, and D. Loss, "Majorana Kramers Pairs in Higher-Order Topological Insulators," Phys. Rev. Lett., vol. 121, no. 19, p. 196801, Nov. 2018, doi: 10.1103/PhysRevLett.121.196801.

[21]  S. Imhof et al., "Topolectrical-circuit realization of topological corner modes," Nat. Phys., vol. 14, no. 9, pp. 925–929, Sep. 2018, doi: 10.1038/s41567-018-0246-1.

[22]  M. Serra-Garcia et al., "Observation of a phononic quadrupole topological insulator," Nature, vol. 555, no. 7696, pp. 342–345, Mar. 2018, doi: 10.1038/nature25156.

[23]  H. Fan, B. Xia, L. Tong, S. Zheng, and D. Yu, "Elastic Higher-Order Topological Insulator with Topologically Protected Corner States," Phys. Rev. Lett., vol. 122, p. 204301, 2043, doi: 10.1103/PhysRevLett.122.204301.

[24]  H. Xue, Y. Yang, F. Gao, Y. Chong, and B. Zhang, "Acoustic higher-order topological insulator on a kagome lattice," Nature Materials, vol. 18, no. 2. Nature Publishing Group, pp. 108–112, Feb. 01, 2019, doi: 10.1038/s41563-018-0251-x.

[25]  X. Ni, M. Weiner, A. Alù, and A. B. Khanikaev, "Observation of higher-order topological acoustic states protected by generalized chiral symmetry," Nat. Mater., vol. 18, no. 2, pp. 113–120, Feb. 2019, doi: 10.1038/s41563-018-0252-9.

[26]  X. Zhang et al., "Second-order topology and multidimensional topological transitions in sonic crystals," Nat. Phys., vol. 15, no. 6, pp. 582–588, Jun. 2019, doi: 10.1038/s41567-019-0472-1.

[27]  X. Zhang et al., "Dimensional hierarchy of higher-order topology in three-dimensional sonic crystals," Nat. Commun., vol. 10, no. 1, pp. 1–10, Dec. 2019, doi: 10.1038/s41467-019-13333-9.

[28]  C. W. Peterson, W. A. Benalcazar, T. L. Hughes, and G. Bahl, "A quantized microwave quadrupole insulator with topologically protected corner states," Nature, vol. 555, no. 7696, pp. 346–350, Mar. 2018, doi: 10.1038/nature25777.

[29]  M. Li et al., "Higher-order topological states in photonic kagome crystals with long-range interactions," Nat. Photonics, vol. 14, no. 2, pp. 89–94, Feb. 2020, doi: 10.1038/s41566-019-0561-9.

[30]  B. Y. Xie et al., "Visualization of Higher-Order Topological Insulating Phases in Two-Dimensional Dielectric Photonic Crystals," Phys. Rev. Lett., vol. 122, no. 23, p. 233903, Jun. 2019, doi: 10.1103/PhysRevLett.122.233903.

[31]  N. H. Burnett, H. A. Baldis, M. C. Richardson, and G. D. Enright, "Harmonic generation in $CO_2$ laser target interaction," Appl. Phys. Lett., vol. 31, no. 3, pp. 172–174, Aug. 1977, doi: 10.1063/1.89628.

[32]  X. D. Chen, W. M. Deng, F. L. Shi, F. L. Zhao, M. Chen, and J. W. Dong, "Direct Observation of Corner States in Second-Order Topological Photonic Crystal Slabs," Phys. Rev. Lett., vol. 122, no. 23, p. 233902, Jun. 2019, doi: 10.1103/PhysRevLett.122.233902.

[33]  B. Xie et al., "Higher-order quantum spin Hall effect in a photonic crystal," Nat. Commun., vol. 11, no. 1, pp. 1–8, Dec. 2020, doi: 10.1038/s41467-020-17593-8.

[34]  M. Proctor et al., "Robustness of topological corner modes in photonic crystals," Phys. Rev. Research, 2, 042038(R), 2020

[35]  J. Noh et al., "Topological protection of photonic mid-gap defect modes," Nat. Photonics, vol. 12, no. 7, pp. 408–415, Jul. 2018, doi: 10.1038/s41566-018-0179-3.

[36]  A. El Hassan, F. K. Kunst, A. Moritz, G. Andler, E. J. Bergholtz, and M. Bourennane, "Corner states of light in photonic waveguides," Nature Photonics, vol. 13, no. 10. Nature Publishing Group, pp. 697–700, Oct. 01, 2019, doi: 10.1038/s41566-019-0519-y.

[37]  S. Mittal, V. V. Orre, G. Zhu, M. A. Gorlach, A. Poddubny, and M. Hafezi, "Photonic quadrupole topological phases," Nature Photonics, vol. 13, no. 10. Nature Publishing Group, pp. 692–696, Oct. 01, 2019, doi: 10.1038/s41566-019-0452-0.

[38]     X. Xie et al., "Cavity Quantum Electrodynamics with Second-Order Topological Corner State," Laser Photon. Rev., vol. 14, no. 8, p. 1900425, Aug. 2020, doi: 10.1002/lpor.201900425.

[39]     W. Zhang et al., "Low-threshold topological nanolasers based on the second-order corner state," Light Sci. Appl., vol. 9, no. 1, pp. 2047–7538, Dec. 2020, doi: 10.1038/s41377-020-00352-

[40]     C. Han, M. Kang, and H. Jeon, "Lasing at Multidimensional Topological States in a Two-Dimensional Photonic Crystal Structure," ACS Photonics, vol. 7, no. 8, pp. 2027–2036, Aug. 2020, doi: 10.1021/acsphotonics.0c00357.

[41]     D. Smirnova, S. Kruk, D. Leykam, E. Melik-Gaykazyan, D. Y. Choi, and Y. Kivshar, "Third-Harmonic Generation in Photonic Topological Metasurfaces," Phys. Rev. Lett., vol. 123, no. 10, p. 103901, Sep. 2019, doi: 10.1103/PhysRevLett.123.103901.

[42]     S. Kruk et al., "Edge States and Topological Phase Transitions in Chains of Dielectric Nanoparticles," Small, vol. 13, no. 11, 2017.

[43]     S. Kruk et al., "Nonlinear light generation in topological nanostructures," Nat. Nanotechnol., vol. 14, pp. 126–130, Dec. 2019, doi: 10.1038/s41565-018-0324-7.

[44]     A. Tripathi et al., "Topological nanophotonics for photoluminescence control," Nanophotonics, vol. 0, no. 0, Sep. 2020, doi: 10.1515/nanoph-2020-0374.

[45]     L. H. Wu and X. Hu, "Scheme for achieving a topological photonic crystal by using dielectric material," Phys. Rev. Lett., vol. 114, no. 22, p. 223901, Jun. 2015, doi: 10.1103/PhysRevLett.114.223901.

[46]     C. W. Peterson, T. Li, W. A. Benalcazar, T. L. Hughes, and G. Bahl, "A fractional corner anomaly reveals higher-order topology," Science, vol. 368, no. 6495, pp. 1114–1118, Jun. 2020, doi: 10.1126/science.aba7604.

[47]     W. A. Benalcazar, T. Li, and T. L. Hughes, "Quantization of fractional corner charge in $C_n$-symmetric higher-order topological crystalline insulators," Phys. Rev. B, vol. 99, no. 24, p. 245151, Jun. 2019, doi: 10.1103/PhysRevB.99.245151.

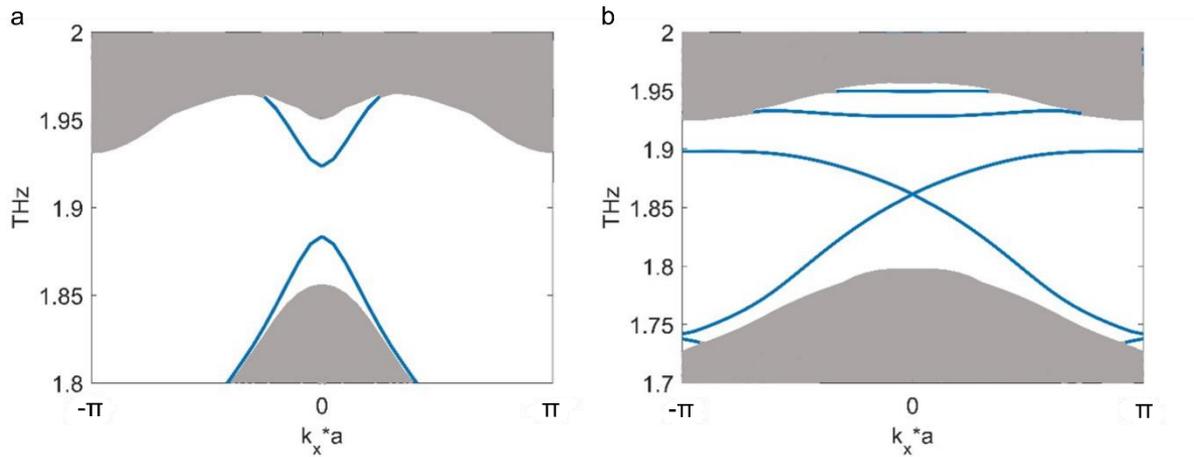

**Figure. S1. Band diagram of arm-chair and zigzag edges.** (a) Arm-chair edge states have mini-band gap in which corner states could reside; (b) zigzag edge states shows diminishing mini-band gap

We will demonstrate below that the gaplessness feature of the zig-zag edge states can be derived from the corner states of arm-chair edges. As is shown in Fig. S2(a) the zig-zag edge can be visualized as combinations of the 120 and 240 degree corners of arm-chair edges, shown as the A and B sites. N is the number of unit cells between the corner sites A and B. As is shown in Figure 3. (d-e) 120 and 240 degree corner states has opposite pairty according to their mirror planes. Hence around the corner states energy, the zig-zag edge can be modeled equivalent to a 1-D tight binding model shown in Fig. S2(b). Note that the hopping strength between adjacent atoms are alternating in signs, due to the parities of the corner modes. This tight binding model construct a 2-band Hamiltonian, whose gap size is soly determined by the difference between the on-site energies $t_1$ and $t_2$, as is shown in Figure. S2(d). Note that in both tight binding model and the full wave simulations the differences of eigen energy of the two corners are diminishingly small, which guarantees the gaplessness of the two edge bands. Figure. S2(c) gives the band diagram with N equals 3. As is shown, the edge band diagram shows even higher similarities to the tight binding model in Figure. S2(d).

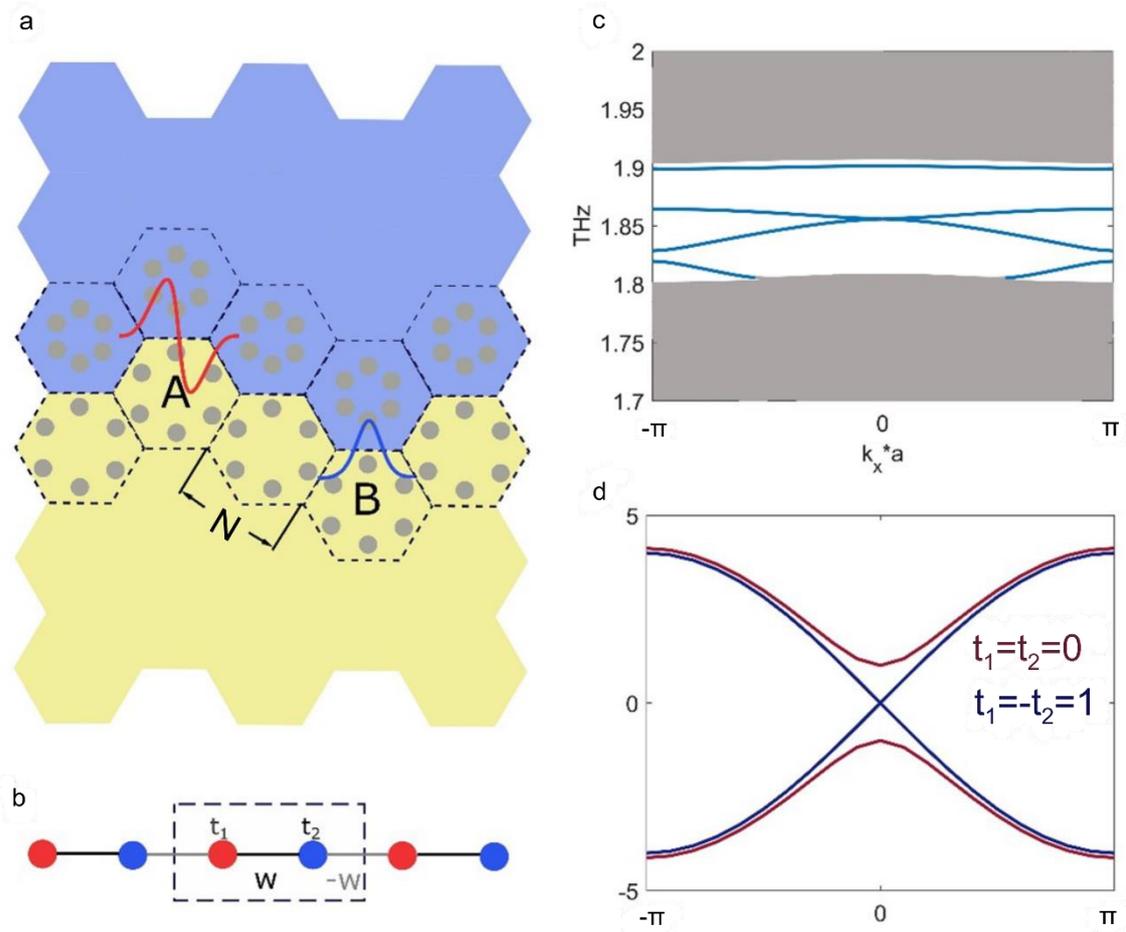

**Figure. S2. Corner states induced band crossing for zig-zag edges.** (a) Schematic of the zig-zag edge. (b) Effective 1D tight binding model for the zig-zag edge Hamiltonian (c) Band diagram for zig-zag edge when the number of unit cells one arm N is 3. (d) Tight binding model's band diagram with W equals 2 and different on-site potentials.

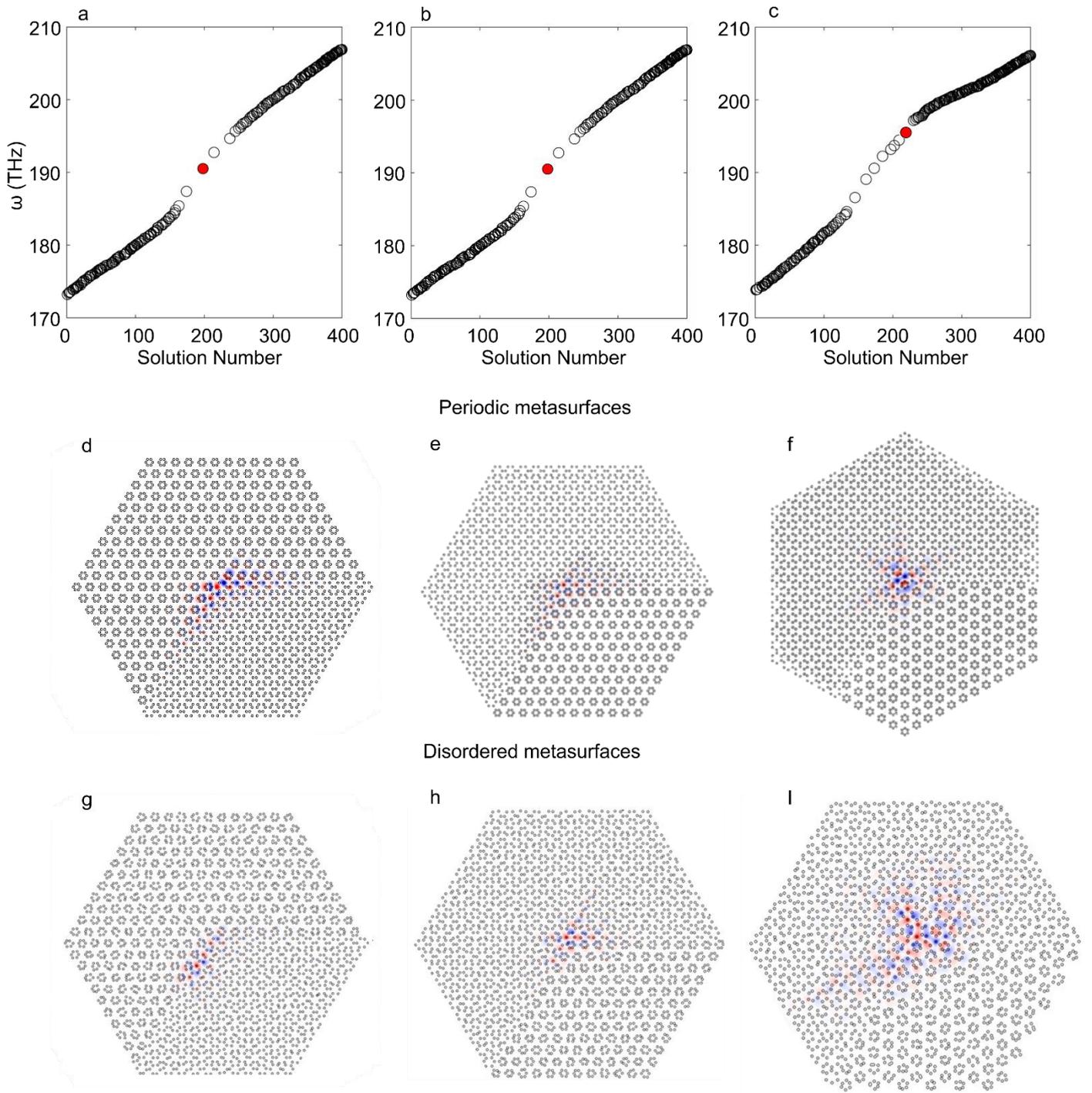

Figure S3. Corner states in ordered and disordered metasurfaces. (a-c) Energy spectra of the systems hosting three types of corner states. (a) Spectrum of a metasurface hosting a 120° angle between arm-chair domain walls featuring a mini-bandgap. (b) 240° angle between arm-chair domain walls. (c) 240° angle between zig-zag domain walls holds trivial corner localizations spectrally near to the bulk state (d-f) Real-space distributions of the electric field around the corners corresponding to red dots marked in a-c. (g-i) Real-space distributions of the electric fields around the corners of the disordered metasurface. Here the perturbation is dislocation of the nanopillars in both x and y directions and satisfy normal distribution with standard deviation of 36 nm.

In Table 1 we provide characteristics of the pump beam and the THG signal used in the spectral measurements in Fig. 4.

**Table 1 characteristics of the pump and the excitation beam.**

| Parameter | Pump wavelength (nm) | Pump pulses repetition rate (MHz) | Pump pulse duration (ps) | Pump average power (mW) | Pump peak power (W) | THG average power (pW) | THG peak power (nW) | Pump coupling | Collection efficiency | Estimated total conversion efficiency [W$^{-2}$] | Estimated total conversion efficiency [dimensionless] |
|---|---|---|---|---|---|---|---|---|---|---|---|
| Symbol | $\lambda$ | $\nu$ | $\tau$ | $P^\omega$ | $P^\omega_p$ | $P^{2\omega}$ | $P^{2\omega}_p$ | $\kappa$ * | $\beta$ ** | $\eta_1$ | $\eta_2$ |
| Equation | | | | | 0.94 $P^\omega/(\nu \tau)$ | | 0.94 $P^{2\omega}/(\sqrt{3} \nu \tau)$ | | | $P^{2\omega}_p / [\beta (\kappa P^\omega_p)^3]$ | $P^{2\omega}_p / [\beta \kappa P^\omega_p]$ |
| Value | 1615 | 5.144 | 5.77 | 119 | 3768.76 | 3.02 | 58.74 | 2.99E-04 | 0.286 | 1.44E-07 | 1.82E-07 |

Pump average power is measured at the position of the sample.

THG average power is estimated from the camera counts integrated over the area of the corner hexamer normalized to the exposure time. Camera quantum efficiency is used to convert counts into Watts. The power measured at the camera detector is then recalculated into the power captured by the collecting objective by taking into account transmittance of all optical components in-between the sample and the detector.

The peak values of the powers are evaluated considering the Gaussian pulse shape, which gives the factor of 0:94.

We assume the third harmonic pulse length differs from the pump pulse by a factor $\sqrt{3}$ which is the approximation valid for a non-resonant bulk material.

We employ a pump coupling coefficient $\kappa$ to account for a portion of power interacting with the corner hexamer (an area at which the THG power was collected). We measure the pump power profile at the sample position with an infra-red camera which approximates well with a Gaussian profile with 30μm FWHM. This corresponds to a beam radius $w_o = \frac{FWHM}{\sqrt{2 \ln 2}} \sim 25.5 \, \mu m$. We evaluate $\kappa$ as the percentage of pump power coming through the top surfaces of six Si pillars forming a hexamer multiplied by 2. The factor 2 comes as a coarse approximation of a surface integral of optical field flux through the resonators [1]. We thus approximate the coupling coefficient as $\kappa = 6 \times 2 \times [1 - \exp(-2r^2/w_o^2)]$ where r is the hexamer pillar's radius.

The collection efficiency is estimated as a portion of $2\pi$ solid angle that falls within the numerical aperture of the collecting objective:

$1 - \mathrm{Sin}[\pi/2 - \mathrm{ArcSin}[NA]]$. We take $2\pi$ solid angle and not full $4\pi$ solid angle as the sample rests on a metallic mirror.

1. K. Koshelev, S. Kruk, E. Melik-Gaykazyan, J.-H. Choi, A. Bogdanov, H.-G. Park, and Y. Kivshar, "Subwavelength dielectric resonators for nonlinear nanophotonics," Science (80-. ). **367**, (2020).